\documentclass{article}

\usepackage[dvips]{graphicx}

\title{Reciprocity and the Emergence of Power Laws in Social Networks}
\author{Michael Schnegg\thanks{Institute of Social Anthropology, University of Cologne, D-50923, Cologne, Germany \texttt{email:michael.schnegg@uni-koeln.de}}}
\date{28.2.2006}

\begin{document}

\maketitle

\begin{abstract}
Research in network science has shown that many naturally occurring and technologically constructed networks are \textit{scale free}, that means a power law degree distribution emerges from a growth model in which each new node attaches to the existing network with a probability proportional to its number of links (=degree). Little is known about whether the same principles of local attachment and global properties apply to societies as well. Empirical evidence from six ethnographic case studies shows that complex social networks have significantly lower scaling exponents $\gamma \sim 1$ than have been assumed in the past. Apparently humans do not only look for the most prominent players to play with. Moreover cooperation in humans is characterized through reciprocity, the tendency to give to those from whom one has received in the past. Both variables -- reciprocity and the scaling exponent -- are negatively correlated ($r=-0.767, sig=0.075$). If we include this effect in simulations of growing networks, degree distributions emerge that are much closer to those empirically observed. While the proportion of nodes with small degrees decreases drastically as we introduce reciprocity, the scaling exponent is more robust and changes only when a relatively large proportion of attachment decisions follow this rule. If social networks are less \textit{scale free} than previously assumed this has far reaching implications for policy makers, public health programs and marketing alike.
\end{abstract}


\section{Introduction}
Networks are a general model to describe complex forms of organization. A network $N=(V, E)$ is defined as a set of vertices $V$ and a set $E$ of unordered pairs of distinct elements of $V$ called links of $V$. Only two networks with $V$ vertices have no structure. One is the network of isolates, where no link is realized. The other is  a completely connected graph in which all possible links are present. Research in network science has shown that some structures are much more likely to occur in the technological and biological world than others~\cite{SG:Mil2004,SG:New2003,SG:Wat98}. Most importantly, many known complex networks share one fundamental property: they are \textit{scale free}~\cite{SG:Bar99}. Each node in a network has $k$ links; $k$ is also called the degree of a node. Networks are \textit{scale free} if the non-cumulative degree distribution $P(k)$ follows a power law: $P(k) \sim 1/k^\gamma $ with an exponent $\gamma$ usually between 2 and 3. \textit{Scale free} networks are dominated by a few hubs, nodes with a very high degree. In contrast the vast majority of nodes have little links. This structure proves very efficient in connecting a random pair of nodes with few links. Moreover \textit{scale free} networks are robust to random failure~\cite{SG:Rek2000,SG:Cohen2000}. In contrast attacks targeted at its hubs can relatively easily disconnect them. To explain the evolution of the \textit{scale free} typology Barab\'asi and Albert have proposed a model that became known as \textit{preferential attachment}. In a network that grows over time, each new node links to the existing structure with a probability that is proportional to the number of links a node already has~\cite{SG:Bar99}. 

Both \textit{preferential attachment} and \textit{scale free} typologies are well established models in the natural sciences ~\cite{SG:New2003,SG:Dor2003,SG:Lus2004,SG:Mon2002}. However, comparatively little is known about whether the same principles  -- local exchanges and the emergent overall typologies -- apply to the social world as well. This is especially astonishing since sociologists, anthropologists, and other social scientists have studied social organization as social networks for almost a century. Consequently no network domain is better documented than the social world. Hundreds of empirical studies describe the social fabric in small groups: fraternities, clubs, organizations, villages and small scale societies~\cite{SG:Was94,SG:Fre2004,SG:Sch98,SG:Wel2001,SG:Sco2000,SG:Whi2004}. Additional survey data shed light on personal networks in some complex societies such as the United States and most European countries. Despite this wealth of empirical information we do not know \textit{whether}, \textit{when}, and \textit{why} the social world follows similar laws. This knowledge would be of enormous value for public health institutions, politics, consumer research and marketing alike. If societies are \textit{scale free} this would make them comparatively easy to control, to influence and to manipulate alike. A characteristic that would play into the hands of those in power.

Besides computer connections, the four most often cited examples of power law distribution in social relationships are citation networks~\cite{SG:Red98}, co-authorships ~\cite{SG:New2004}, co-acting relations~\cite{SG:Bar99}, and sexual relationships~\cite{SG:Lil2001}. It is  difficult to treat any of these relations as an adequate proxy for the larger construct of social relations in a complex social world. Co-acting, citing and even co-authoring and sexual intercourse are only relatively selected aspects of the spectrum of social interactions humans engage in. While combinations of these networks would make up a better indicator of the latent variable we do not know whether they are correlated or not. Often they will not be. The best scientist is unlikely to be the most popular actor at the same time. Not to speak of lovers. Social activities are not only restricted by talent but also by time. To have many co-authors produces opportunity costs. Some evidence exists that there is an upper bound of social networks size that is related to neocortex size~\cite{SG:Hil2003}. However, if different social relations are only loosely correlated, or not at all the overall pattern of the multi-relational network would override its effects and make the aggregate significantly less \textit{scale free} than any of its parts. 

From the perspective of a social scientist a better proxy for the social structure may be sets of interactions that link across different societal domains. Friendship could be an indicator of these relationships. For a more in-depth understanding the typology of the networks emerging from more encompassing, time consuming, and emotionally laden relationships needs to be analyzed. Drawing on common sense and empirical observations in different societies we question the fact that 80 percent of all friends really belong to a happy few. But this is what a power law in friendship ties translates to. 

\section{Method and Data}
Empirical data on incoming social ties is needed in order to test whether social networks of the type we are interested in are \textit{scale free} or not. This restricts the availability of data sets significantly. Personal network surveys, the most common and effective way to collect social network data in complex societies, focus on outgoing ties, the people an individual asks for social support in a given situation. To overcome these constraints we consider only complete networks in this comparison. Many of the data sets presented in Table 1 were collected to understand the significance of social networks and social capital as a coping strategy to reduce household vulnerability in the extremely insecure ecological, economic and political environment in Sub-Saharan Africa. All participating ethnographers have tried to capture the most salient aspects of social relations in the given society. Most of the studies were conducted in rural villages. The collection of social network data is often easier in small communities than in large cities. The social systems are still comparably bounded and their boundaries are easier to define. Table 1 gives the summary network statistics for six different societies. 

\begin{table}[t]
\centering
\begin{tabular}{|l|c|c|c|c|c|c|c|}
\hline
Society & N & dir. links & undir. links & Recip. & $\gamma$ & max $\gamma$ \\
\hline
Tlaxcala  & 142  & 249 & 7 & 5.3 & 1.67 & 2.19 \\
Herero & 41 & 67 & 18 & 34.9 & 0.77& 1.67  \\
Ju/'hoansi  & 73 & 123 & 58 & 48.5 & 1.05 & 1.93 \\
Pokot  & 37 & 182 & 149 & 62.1 & -0.25 & 1.53 \\
Damara & 62 & 74 & 66 & 66.1 & 0.75 & 1.94  \\
US Fraternity & 58 & 207 & 106 & 50.6 & 0.35 & 2.02 \\
\hline

\end{tabular}
\caption{Network statistics of 6 social networks, mostly small-scale societies from Africa}
\end{table}

\begin{figure}[t]
\centering
\includegraphics[angle=0,scale=.60]{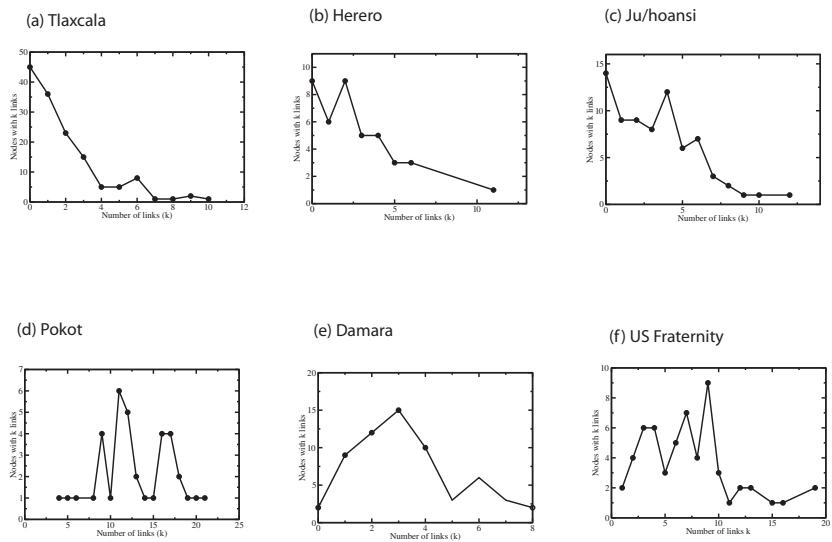}
\caption{Degree distributions of the 6 social networks: Number of nodes with $k$ links each, versus $k$.}
\end{figure}

\textbf{Tlaxcala} is a state in Mexico. Apart from kinship \textit{compadrazgo} (ritual kinship) is the second most important building block of social order. \textit{Compadrazgos} developed out of the godparenthood relations. However \textit{compadrazgos} cover a much wider range of rites of passage, including first hair cut, graduation, marriage, and inauguration of new possessions. The \textit{compadre} serves as a ritual sponsor in these events. Tlaxcala is an industrialized state in which the textile industry is the most important sector. The data were collected in 1975 by interviewing 142 couples about all the \textit{compadrazgos} relationships they had engaged in over the course of their lives ~\cite{SG:Whi2002}.

The \textbf{Herero} are pastoralists of central Namibia. Herero social organization is dominated by the rare double decent kinship system in which an individual belongs both to the mothers' and fathers' unilineal kinship group. The data analyzed here were collected in 1993 by Ute Stahl (University of Cologne, Germany) and describe support relations among a set of pastoral households. 

The \textbf{Ju/'hoansi} are hunter and gatherers of Namibia and Botswana. At the time of the data collection in the 1970ies the Ju/'hoansi lived in relatively small and mobile bands. In addition to kinship, exchanges of gifts (\textit{hxaro}) are fundamental social relationships. \textit{Hxaro} relations guarantee access to water holes and other resources in distinct areas. The data analyzed here describe these transactions for a set of individuals ~\cite{SG:Wie82,SG:Wie2002}.

The \textbf{Pokot} are pastoralists in North-west Kenya. At the time of the data collection in 1987 the society was characterized by a strong egalitarian ethos. The main source of wealth and prestige are cattle. Pokot society is structured according to age sets and the unilinial kinship system. The elders have a strong say in most social affairs. The networks analyzed here describe transactions between households in one Pokot neighborhood~\cite{SG:Bol2000}.

The \textbf{Damara} are agro-pastoralists of central Namibia. Their social structure is largely dominated by exchanges that anthropologists have called demand sharing. Demand sharing refers to sharing that is initiated through a verbal demand. In Khoekhoegowab, the language of the Damara people, this demand is expressed as \textit{"Au te re X!"} (\textit{Give me some X!}). The person being asked is morally obliged to give some share of what s/he has. A common strategy to avoid exploitation is to hide the true amount of an asset and to only give a "fair" share of what is visible to the demanding individual. The social linkages analyzed here are transactions, including food and other goods needed for daily survival. They were collected in 2004 by the author and Julia Pauli over a period of 10 days by interviewing each of the 62 households at the end of the day about all in- and outgoing transactions that had taken place during the last 24 hours.

The \textbf{fraternity data} record interactions and perceived interactions among all members of a fraternity at West Virginia college. The data were collected to analyze the relationship between perception and behavior in social networks ~\cite{SG:Ber80}. 

Table 1 shows that the effective scaling exponent from a least-squares fit does not reach a value higher than 1.6 in any of the six cases discussed. In addition to the observed values $\gamma$, Table 1 also gives the highest possible scaling exponent, max $\gamma$, for the given size and density of the network. This value was computed using a Barab\'asi-Albert (BA) simulation in which the number of links was fixed to the mean of the observed network. A program for the simulation is described in Stauffer et at. \cite{SG:Sta2006}. The comparison shows that low $\gamma$ values are not an artifact of small sample size. Given samples would have much higher scaling exponents if \textit{preferential attachment} is the only rule of choice.

The six corresponding distributions are shown in Figure 1. Tlaxcala is the only case that shows relatively clear characteristics of a \textit{scale free}  typology. There is a straightforward explanation to this: the ritual sponsorship (\textit{compadrazgo}) is very costly. Only a very few people can afford to pay for the many \textit{fiestas} in which the relationships are celebrated. The other data sets deviate stronger from a pure power law model. In general, distributions fall into two categories: scale free like networks in the top row of Figure 1 and random networks in the bottom row. In none of the later cases the mode of the distribution is the smallest degree. Now what distinguishes these networks? If we look at the fourth column of Table 1 we see one parallel. The networks at the bottom are networks with high levels of reciprocity. Reciprocity is defined as the fraction in the overall link distribution of links such that A is tied to B if B is tied to A, the non-reciprocal links are directed, from A to B or from B to A. Networks at the top have lower reciprocity. Reciprocity is defined as the percentage of reciprocal ties in the overall link distribution. These two columns -- reciprocity and $\gamma$ -- show a clear relationship: As reciprocity increases, the scaling exponent decreases (Pearson $r=-0.767$). The relationship is significant at the level 0.075.

\section{Beyond Preferential Attachment: Reciprocity}
Preferential Attachment, linking with central players, is a plausible rule to explain the emergence of social order among humans. When the physicists Barab\'asi and Albert coined the term in 1999 they were not aware that this rule had been acknowledged in the social sciences for quite some time. In 1953 Moreno speculated on a general sociological law behind the power law distribution he observed in friendship choices~\cite{SG:Mor53} and Merton called the rule in his sociology of science "the Matthew effect": "For unto every one that hath shall be given, and he shall have abundance"~\cite{SG:Mer68}. Most explicitly De Solla Price found scientific citations to follow a power law with an exponent in the range of 2.5-3.0 and called the law behind this formation Cumulative Advantage~\cite{SG:Pri76}. Even though the idea that links in networks are not equally distributed has been debated in the past, Barab\'asi and Albert recognized and established the exact relationship between local attachment (exchange) rules and emerging network typologies~\cite{SG:Bar99}. Over the last years others have refined their model to account for directed ties and more specific local circumstances~\cite{SG:Dor2003,SG:Puj2005,SG:Jin2001,SG:Pen2002,SG:Gib2005}. Most importantly for the social sciences Caldarelli et al. showed that social ties reconstructed from five email folders indicate the importance of preferential exchange~\cite{SG:Cal2004}. The likelihood of interactions depends on the history of previous interactions between any two individuals. The authors found their email traffic was concentrated on a few communication partners while for most addresses only a few messages were exchanged. They were the first to acknowledge the importance of memory this debate. 

Over the last 20 years a significant body of research in anthropology, economics and sociology has shown that humans are not only forward looking utility maximizers. \textit{Homo reciprocans} may be a much more adequate model of man than its forefather \textit{homo oeconomicus}. The anthropologists Mauss and Levi-Strauss recognized the importance of reciprocity, the giving and taking of gifts, goods, and services as one of the primary principles of social organization~\cite{SG:Mau25,SG:Str49}. Evolutionary theory recognizes two pathways to cooperation between organisms: kin-selection and reciprocal altruism. Whereas kin-selection can explain cooperation between kin, evolutionary theorists, including Darwin, have long stumbled to explain, why organisms deliver benefits to unrelated organisms. Trivers' concept of reciprocal altruism and its game theoretic implementations by Hamilton and Axelrod are the most convincing solutions to this problem~\cite{SG:Axe81,SG:Tri71}. The fitness of organisms can be improved through long run mutual interactions as long-term as the benefits delivered by each are greater than the costs incurred. The principle is equivalent to the economic principle of mutual gains in trade. Reciprocity outside the close kinship domain is not only present with almost all apes but also with vampire bats and other mammals~\cite{SG:Wil84}.
Most recent intercultural work in experimental economics demonstrates that fairness and reciprocity may well be universal in humans. In the Ultimatum Game (UG) player A is given a  fixed sum of money and asked to allocate the money between himself and a player B. The identity of player B is unknown to him. Player B is then informed about the allocation and has the option to accept or to reject the offer. If s/he rejects, both go home with nothing.  The UG was recently played in 15 different small scale societies. In all societies players allocated appreciable amounts of money to their counterparts and rejected offers close to zero. A utility maximizing individual would have done something different. It would have taken whatever was on the table~\cite{SG:Hen2004}. Largely different research traditions underline the importance of reciprocity as a fundamental trait of cooperation. But would a network of individuals that have limited resources to maintain relationships and try to maintain a certain level of reciprocity look the same as a \textit{scale free} network? Simulations allow the testing of what effects reciprocity as a local rule of attachment has on the network structure as a whole.

\section{The Simulation Model}
The aim of the model is to test what happens when people in a network start playing fair instead of only looking after important hubs when choosing new exchange partners. The simulation starts with a loosely connected Erd\"os-R\'enyi-type network in which each node picks two partners at random and gives them something (degree=2). In the second step it iterates through 10 rounds of exchanges for each player. The starting condition is different from the seed usually defined in the BA model. In the BA model the starting configuration is a small, fully connected graph. Since we want to allow all nodes to choose partners according to previous exchanges as well, we must define a \textit{status quo} before running the actual simulation. The easiest \textit{status quo} is a random graph. Throughout all 10 rounds of the simulation we allow all 10,000 vertices to select partners to give something to according to one of the following two rules: (1) pick your partner with a probability proportional to its number of incoming links  (\textit{preferential attachment}) or (2) randomly pick one of the players who gave you something in the past (\textit{reciprocity}). The first choice is made with probability $1-P$; otherwise we take the second choice, randomly at each iteration. The resulting degree distributions are stored and averaged over 10 simulation runs. 

\begin{figure}[h]
\centering
\includegraphics[angle=0,scale=.25]{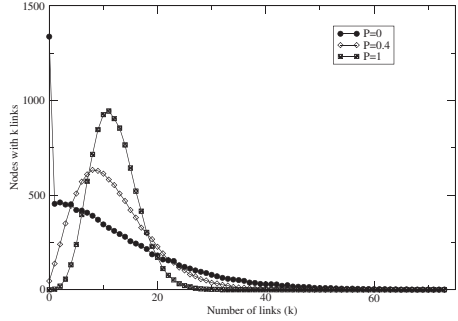}
\caption{Degree distributions for different models. Results are averages over 10 simulation runs.}
\end{figure}

Figure 2 shows the degree distributions with varying probabilities $P$, that actors play reciprocal. The two extremes, $P=0$ (full BA) and $P=1$ (full reciprocity), are well known: A power law distribution on the one hand and a Poisson distribution on the other. Effective values of  $\gamma$ between those two extremes are combinations of the two rules. Adding only small percentages of the reciprocity rule to the exchange system alters its structure. Table 2 summarizes the results and shows the transformation from a \textit{scale free} to a Gaussian typology. 

\begin{table}[h]
\centering
\begin{tabular}{|c|c|c|}
\hline
0=BA,1=Reciprocity &  $\gamma$ & First Quartile \\
\hline
0.0 & 1.98 & 3\\
0.1 & 1.88 & 4\\
0.2 & 1.92 & 5 \\
0.3 & 1.68 & 6\\
0.4 & 1.78 & 7\\
0.5 & 1.42 & 7\\
0.6 & 1.36 & 8\\
0.7 & 1.28 & 8\\
0.8 & 1.24  & 8 \\
0.9 & 1.15  & 9 \\
1.0 & 0.71  & 9 \\
\hline
\end{tabular}
\caption{Averaged network statistics of 10 simulation runs ($N=10,000$)}
\label{tab:}
\end{table}

The scaling exponent $\gamma$ proves to be comparatively robust. The long tail of the distribution remains even if we add 30 or 40 percent of reciprocity to the exchange system. The second column gives the value where the first 25 percent of the distribution ends. This number is called the first quartile. We use it to describe the behavior at the left end of the distribution. In the full BA model 25 percent of the nodes have 3 links or less. In the simulation this proportion of nodes with very few links decreases much faster than the scaling exponent.  This matches the empirical distributions described above. In the true BA model the mode (maximum) of the distribution is the smallest number of links. Four of our empirical examples differ from this. The mode is not the smallest value. As the simulation shows this effect can be caused by reciprocity. 

Our observation that many social networks show power law distributions with exponents lower than 2 matches the simulation. Blending reciprocity and memory into transactions reproduces networks much better correlated to the social world than the BA model alone. 

\section{Implications}
The empirical evidence from six different societies shows marked variations from fully \textit{scale free} networks. Drawing on theoretical work in both the social sciences and evolutionary psychology, we included reciprocity as an additional exchange rule to account for different network typologies and scaling exponents lower than 2. The implications of our findings are far reaching. Information, opinions, and viruses spread very differently in networks with different typologies. 
\newline
\newline
\textbf{Acknowledgment:} Dietrich Stauffer taught me how to simulate networks. He also developed the algorithm used in the simulation. For this assistance, his careful reading of the manuscript and his support I am very thankful. Hopefully one day I can reciprocate. I thank  Russ Bernard, Michael Bollig, Julia Pauli, Ute Stahl, Doug White and Polly Wiessner who have contributed data for this comparison. Discussions with Patrick Heady, Julia Pauli and Michael Bollig about the fundamental importance of reciprocity in small scale societies stimulated this work. Doug White gave valuable comments to improve the manuscript. The research was funded by the German National Science Foundation (DFG) as part of the SFB 389 ACACIA.

\end{document}